\documentclass[fleqn,usenatbib]{mnras}

\usepackage{ulem}
\usepackage[T1]{fontenc}

\DeclareRobustCommand{\VAN}[3]{#2}
\let\VANthebibliography\thebibliography
\def\thebibliography{\DeclareRobustCommand{\VAN}[3]{##3}\VANthebibliography}

\usepackage{graphicx}	
\usepackage{amsmath}	
\usepackage{amssymb}	
\usepackage[dvipsnames]{xcolor}

\usepackage{newtxtext,newtxmath}

\newcommand{\ie}{i.e.,~}
\newcommand{\eg}{e.g.,~}

\begin{document}



\title[GRMHD simulations of accreting neutron stars I]{GRMHD simulations
  of accreting neutron stars I: nonrotating dipoles}

\author[S. \c{C}\i{}k\i{}nto\u{g}lu et al.]{
Sercan \c{C}\i{}k\i{}nto\u{g}lu,$^{1,2}$\thanks{E-mail: cikintoglus@itu.edu.tr}
K. Yavuz Ek\c{s}i,$^{1}$
Luciano Rezzolla$^{2,3,4}$
\\
$^{1}$Istanbul Technical University, Faculty of Science and Letters,
Physics Engineering Department, 34469, Istanbul, Turkey\\
$^{2}$Institut f\"ur Theoretische Physik, Goethe-Universit\"at,
Max-von-Laue-Str. 1, 60438 Frankfurt am Main, Germany\\
$^{3}$Frankfurt Institute for Advanced Studies, Ruth-Moufang-Str. 1,
60438 Frankfurt am Main, Germany\\
$^{4}$School of Mathematics, Trinity College, Dublin 2, Ireland }

\date{Accepted XXX. Received YYY; in original form ZZZ}

\pubyear{2021}

\label{firstpage}
\pagerange{\pageref{firstpage}--\pageref{lastpage}}
\maketitle

\begin{abstract}
We study the general-relativistic dynamics of matter being accreted onto
and ejected by a magnetised and nonrotating neutron star. The dynamics is
followed in the framework of fully general relativistic
magnetohydrodynamics (GRMHD) within the ideal-MHD limit and in two
spatial dimensions. More specifically, making use of the numerical code
\texttt{BHAC}, we follow the evolution of a geometrically thick matter
torus driven into accretion by the development of a magnetorotational
instability. By making use of a number of simulations in which we vary
the strength of the stellar dipolar magnetic field, we can determine
self-consistently the location of the magnetospheric (or Alfv\'en) radius
$r_{\rm msph}$ and study how it depends on the magnetic moment $\mu$ and
on the accretion rate. Overall, we recover the analytic Newtonian scaling
relation, \ie $r_{\rm msph} \propto B^{4/7}$, but also find that the
dependence on the accretion rate is very weak. Furthermore, we find that
the material torque correlates linearly with the mass-accretion rate,
although both of them exhibit rapid fluctuations. Interestingly, the
total torque fluctuates drastically in strong magnetic field simulations
and these unsteady torques observed in the simulations could be
associated with the spin fluctuations observed in X-ray pulsars.
\end{abstract}




\section{Introduction}

Since the discovery of the first pulsating X-ray sources Centaurus X-3
\citep{Giacconi1971,Schreier1972} and Hercules X-1 \citep{Tananbaum1972},
accretion of matter transferred from a binary companion onto a neutron
star \citep{Pringle72,Davidson1973,Lamb1973} is known to power many
classes of X-ray sources from the nuclear- and accretion-powered
millisecond X-ray pulsars \citep{Wijnands1998} in low-mass X-ray binaries
to pulsating ultra-luminous X-ray sources \citep{Bachetti2014} in
high-mass X-ray binaries. In the former case, mass transfer occurs by
Roche-lobe overflow and proceeds by accretion from a  viscously
evolving disc around the compact object.

Furthermore, if the dipolar magnetic field of the neutron star is
sufficiently strong, the inner portions of the disc are disrupted and the
flow is funnelled onto the star along the magnetic field lines. The
magnetospheric (or Alfv\'en) radius of the disc is located at the
distance where the material stresses are balanced by the magnetic
stresses \citep{Ghosh1977, Ghosh1978, Ghosh1979a, Ghosh1979, Wang1987,
  Spruit1990, Campbell1992, Wang1996} (see the detailed discussion in
Sect. \ref{sec:rmsph}).

Some of the early models assumed that the disc, due to the high
conductivity of its ionised plasma, would exhibit a diamagnetic behaviour
\ie expel the stellar magnetic field lines except for the innermost
boundary where matter is coupled to the field lines
\citep{Ichimaru1978,Scharlemann1978,Aly1980}. This mode of interaction
would then lead only to a spin-up of the neutron star via
accretion. However, the discovery of X-ray pulsars that spin-down in the
accretion stage motivated the ``magnetically threaded disc'' model
\citep{Ghosh1979a,Ghosh1979,Kaburaki1986}, where the stellar field lines
are allowed to penetrate the disc over a broad region via several
instabilities. This model assumes that the poloidal stellar field couples
to the toroidal field generated from the angular-velocity difference
between the star $\Omega_*$ and that of the matter in the disc
$\Omega(r)$. After defining the magnetic pitch factor as the ratio
between the toroidal ($B_{\rm tor}$) and poloidal ($B_{\rm pol}$)
components,\footnote{In earlier analytical studies where the disc was
assumed to be geometrically thin, the radial magnetic-field component in
the disc has been neglected, so that $B_{\rm pol} = B^{\theta}$. However,
in geometrically thick discs the radial component can be important and
our numerical simulations clearly show that not only the radial component
is nonzero, but also that it can be the locally dominant component.}  \ie
$\gamma_\phi := B_{\rm tor}/B_{\rm pol}$, the magnetically threaded disc
model implies that $\gamma_{\phi} \propto (\Omega_* -
\Omega(r))/\Omega_{_{\rm K}}(r)$ -- where $\Omega_{_{\rm K}}$ is the
Keplerian angular velocity -- so that a negative magnetic torque
proportional to $\gamma_{\phi}$ is exerted onto the star by the
magnetic-field lines penetrating the disc beyond the corotation radius,
which is where the disc rotates at the same angular velocity of the star.

The magnetic pitch factor is an important ingredient of the magnetically
threaded model and the dependence of the pitch factor on the velocity
difference -- as well as on the processes limiting the growth of the
toroidal magnetic field in regions where the velocity difference is large
-- has been explored in the literature \citep{Wang1987, Campbell1992,
  Wang1995, Uzdensky2002}. In particular, it has been pointed out that
the model of \citet{Ghosh1979a} tends to produce excessively large
toroidal magnetic fields at large distances from the corotation axis, so
large that the resulting magnetic pressure would disrupt the disc beyond
the corotation radius \citep{Wang1987}, unless some magnetic-field
diffusive mechanism is invoked in the disc \citep{Campbell1992,
  Wang1995}.

More recently, steady-state configurations of the toroidal magnetic field
was studied by \citet{Naso2011}, who solved the
full induction equation and observed that the profile of the toroidal
magnetic field is rather different from that proposed in previous
analytical models of \citet{Wang1987} and \cite{Campbell1992}. We note
that the spin-up timescale of X-ray pulsars, $\tau :=
\Omega/|\dot{\Omega}| \lesssim 10^4\,{\rm yr}$, is much less than their
lifetime as accretors ($\tau_{\rm X} \gtrsim 10^6\,{\rm yr}$). This
implies that the long-term spin-up trends observed from the first
discovered X-ray pulsars do not represent a steady-state behaviour
\citep{Elsner1980}. Indeed, it was soon observed that some accreting
pulsars, such as GX~1+4 \citep{Makishima1988}, 4U~1626-67
\citep{Chakrabarty1997,Camero2010} and 4U~1907+09 \citep{Inam2009}, show
torque reversals with secular spin-up or spin-down on sufficiently long
timescales. Furthermore, observations of accreting pulsars with Burst and
Transient Source Experiment (BATSE) on the Compton Gamma Ray Observatory
showed that all accreting pulsars, independently of whether they are
disc-fed or wind-fed, exhibit stochastic variations in their spin
frequencies and luminosities \citep{Bildsten97}. BATSE observations also
showed that sources exhibiting secular spin-up and spin-down trends also
have stochastic variations in their spin frequencies. 

In view of these considerations, over the last decade a number of
numerical simulations solving the full set of the magnetohydrodynamics
(MHD) equations, either in two or three spatial dimensions, have explored
the complex behaviour of the interaction between a magnetised star and an
accretion disc. While these simulations are also subject to
simplifications and approximations, they are expected to provide a more
accurate description of the complex non-stationary behaviour that is
expected in accreting plasmas and that is suggested by the
observations. Indeed, such simulations have revealed that the accretion
flow onto a compact object from a magnetised plasma is intrinsically
accompanied by the development in the disc of a magnetorotational
instability \citep[MRI;][]{Velikhov1959, Chandrasekhar1960} that leads to
a turbulent flow with a steady behaviour accompanied by fluctuations in
all the physical quantities \citep{Balbus1991}.

Simulations providing useful insight for our understanding of the
disc-magnetosphere interaction have been carried out initially within a
Newtonian description of gravity \citep{Hayashi1996, Miller1997,
  Romanova2002, Long2005, Bessolaz2008, Romanova2008, Kulkarni2008,
  Romanova2011, Romanova2012, Kulkarni2013, Romanova2015,
  Ireland2022}. However, general-relativistic effects might play an
important role in accretion onto neutron stars, especially in the case of
weakly magnetised neutron stars in low-mass X-ray binaries, where the
disc can extend closer to the star \citep{Psaltis1999b}. Hence,
general-relativistic MHD (GRMHD) simulations have been performed recently
to explore ultra-luminous X-ray sources \citep{Takahashi2017b,
  Takahashi2018, Abarca2018, Abarca2021}, and accreting millisecond X-ray
pulsars for dipole magnetic field \citep{Parfrey2017} and multipole
magnetic field geometries \citep{Das2022}.

We here report GRMHD simulations of accretion onto a magnetised
nonrotating neutron star from a MRI driven accretion torus. In
particular, we examine general relativistic effects on the
disc-magnetosphere interaction, specifically the modifications on the
magnetospheric radius. To this end, we perform simulations for ten
different magnetisations of the neutron star and study how the
magnetospheric radius depends on the stellar magnetic-field strength and
on the properties of the accretion flow. We also investigate the
properties of the magnetic pitch factor in the presence of MRI-induced
turbulent fields.

The structure of the paper is as follows. The numerical setup and the
initial conditions of our simulations are introduced in
Sec.~\ref{sec:NuSe}, while the results of the simulations are presented
in Sec.~\ref{sec:Re}. Finally, Sec.~\ref{sec:Dis} collects the discussion
of the results and our conclusions. Hereafter, and unless indicated
otherwise, we adopt geometrised units where $G=c=1$, with $G$ and $c$
being the gravitational constant and the speed of light, respectively.

\section{Numerical Setup}
\label{sec:NuSe}

In our simulations we employ \texttt{BHAC} \citep{Porth2017,Olivares2019}
to solve numerically the full set of the GRMHD equations over the fixed
spacetime of a nonrotating star
\begin{align}
\nabla_\mu J^{\mu} =&\, 0\,, \\
\nabla_\mu T^{\mu\nu} =&\, 0\,, \\
\nabla_\mu {}^*\!F^{\mu\nu} =&\, 0\,,
\end{align}
where $J^{\mu}$ is the rest-mass current, $T^{\mu\nu}$ is the total
energy-momentum tensor \citep{Rezzolla_book:2013} and $F^{\mu\nu}$ the
Faraday tensor. More specifically, the explicit expression for rest-mass
current and the energy-momentum tensor of a magnetised perfect fluid are
given by
\begin{align}
  J^{\mu} &= \rho u^\mu\,,&\\
  T^{\mu\nu}&=\left(\rho h + b^2\right)u^\mu
  u^\nu + \left(p+\frac{b^2}{2}\right)g^{\mu\nu}-b^\mu b^\nu\,,&
\end{align} 
where $h$, $\rho$, $p$, $u^\mu$, $g_{\mu\nu}$, $b^\mu$ are, respectively,
the specific enthalpy, the rest-mass density, the pressure, the fluid
four-velocity, the metric tensor, and the magnetic field measured in the
fluid frame, so that $b^2:=b^\mu b_\mu$. In the ideal-MHD limit
considered here, the electric field in the comoving frame is zero, \ie
$F^{\mu\nu}u_{\mu}=0$, and the dual of the Faraday tensor is given by
\begin{equation}
{}^*\!F^{\mu\nu} = u^\mu b^\nu - u^\nu b^\mu\,.
\end{equation}
Additional quantities used hereafter are: the Eulerian three-velocity,
$v^i:=u^i/\Gamma + \beta^i/\alpha$, where $\Gamma:= (1-v^iv_i)^{1/2}$ is
the Lorentz factor -- $\alpha$ and $\beta^i$ are, respectively, the lapse
function and the components of the shift vector in a 3+1 decomposition of
the four-metric -- and the Eulerian magnetic field,
$B^i:={}^*\!F^{i\nu}n_\nu$, where $n_\mu = \left(-1/\sqrt{-g^{tt}}, 0, 0,
0\right)$ represents the one-form associated to an Eulerian observer
\citep{Rezzolla_book:2013}.

We initialise the simulations with a Fishbone \& Moncrief (FM) torus
\citep{Fishbone76}, where its inner edge and the rest-mass density
maximum are located at $r_{\mathrm{in}}=200\,M_{\odot}$ and
$r_{\mathrm{max}}=260\,M_{\odot}$, respectively. At such a distance, the
dipole magnetic field of the star is too weak to spoil the equilibrium of
the torus. The system of GRMHD equations is closed with an equation of
state with of an ideal fluid $p=\rho\epsilon\left(\gamma-1\right)$, where
$\epsilon$ is the specific internal energy and we assume an adiabatic
index $\gamma=5/3$ as for a completely degenerate non-relativistic
electron fluid \citep{Rezzolla_book:2013}.

On the other hand, the mass and the radius of the neutron star are taken
to be $M=1.4\,M_{\odot}$ and $R=14\,\mathrm{km}$,
respectively. Furthermore, because the mass of the accreting fluid is
much smaller than that of the nonrotating neutron star, the spacetime
outside the star is described by the Schwarzschild metric
\begin{equation}
d s^2=-\left(1-\frac{2M}{r}\right)\, d
t^2+\left(1-\frac{2M}{r}\right)^{-1}\,d r^2+r^2 \,
d\theta^2+r^2\sin^2\theta \, d\phi^2\,.
\end{equation}
The interior spacetime, on the other hand, does not need to be specified
in our implementation since suitable boundary conditions will be imposed
at the stellar surface (see discussion below).

Since the star is magnetised, a proper general-relativistic solution of
the Maxwell equations for the external dipolar magnetic field needs to be
specified. We here employ the analytic expressions derived by
\citet{Wasserman1983} and \citet{Rezzolla2001}, and consider ten
different strengths of the stellar magnetic field. In particular, we
specify the dipolar magnetic moment $\mu$ such that the maximum
magnetisation parameter is
\begin{equation}
\sigma_{\max} := \frac{b_{\mathrm{pole}}^2}{\rho_{\max}}
=0.08433\times\left\lbrace
1, 4, 9, 16, 25, 36, 49, 64, 81, 100\right\rbrace\,,
\end{equation}
where $b_{\mathrm{pole}}^2$ is the strength of the magnetic field at the
pole of the star and $\rho_{\max}$ is the maximum rest-mass density of
the torus. In addition to the dipolar field of the star, a weak poloidal
magnetic field is seeded in the torus via the toroidal component of a
vector potential $A_\phi\propto\max\left(\rho/\rho_{\max}-0.2,0\right)$.
The maximum of the rest-mass density is also used to introduce the
following dimensionless dynamical quantities
\begin{equation}
\tilde{\rho} := \frac{\rho}{\rho_{\max}}\,, \qquad
\tilde{p} := \frac{p}{\rho_{\max}}\,, \qquad
\tilde{B} := \frac{B}{\sqrt{\rho_{\max}}}\,,
\end{equation}
so that all of our results can be scaled simply in terms of the choice
for $\rho_{\max}$ (to keep the notation compact, we will omit the tildes
on the relevant quantities). 

Furthermore, since we do not include radiation hydrodynamics in our
simulations [see instead \citet{Takahashi2017b,Abarca2018} where this is
  considered for accretion onto a non-magnetised star or
  \citet{Takahashi2018} for a magnetised star], our results are viable
only in the sub-Eddington regime. Hence, in order to keep the
mass-accretion rate lower than the Eddington limit and, at the same time,
have the stellar magnetic field sufficiently strong so that the
magnetospheric radius is larger than the stellar radius, we have set
$\rho_{\max} = 0.001\,\mathrm{g~cm^{-3}}$. Hereafter, we present our
results in units of
\begin{align}
  &B_9:= \left(\frac{B}{10^9\,\mathrm{G}}\right)
  \left(\frac{10^{-3}\,\mathrm{g~cm^{-3}}}{\rho_{\max}}\right)^{1/2}\,, &\\
  &\mu_{27}:= \left(\frac{\mu}{10^{27}\,\mathrm{G~cm^3}}\right)
  \left(\frac{10^{-3}\,\mathrm{g~cm^{-3}}}{\rho_{\max}}\right)^{1/2}\,, &\\
  &\dot{m}:=\left(\frac{\dot{M}}{\eta\dot{M}_{\mathrm{Edd}}}\right)
 \left(\frac{10^{-3}\,\mathrm{g~cm^{-3}}}{\rho_{\max}}\right)\,,& \\
  &\dot{j} := \left(\frac{\dot{J}}{10^{34}\,\mathrm{g~cm^2~s^{-2}}}\right)
  \left(\frac{10^{-3}\,\mathrm{g~cm^{-3}}}{\rho_{\max}}\right)\,,&
\label{eq:units}
\end{align} 
where $\dot{M}_\mathrm{Edd}:= 2.2 \times 10^{-9} (M/M_{\odot})\,
M_{\odot}~{\rm yr}^{-1}$ is the Eddington mass-accretion rate and
$\eta:=(GM/Rc^2)\simeq 0.15$ is the efficiency coefficient. We note that
magnetic-field values of the order of $\lesssim 10^{10}\,{\rm G}$ are
necessary to avoid regions of excessive magnetizations that cannot be
handled within a GRMHD solution; similar values of the magnetic field
have been used in the literature \citep{Romanova2011, Romanova2012,
  Romanova2015}, but see also \citet{Parfrey2017} and \citet{Das2022}.

The stellar dipolar magnetic field and poloidal field of the torus are
set to be anti-parallel as this configuration has been shown to lead to a
smoother accretion of matter \citep{Romanova2011}, essentially because
the larger amount of reconnection taking place at the magnetospheric
boundary releases larger amounts of magnetic energy and favours a more
copious accretion; simulations in which the magnetic fields are parallel
will be presented in a subsequent work. The magnitude of the poloidal
magnetic field in the torus is set such that the plasma parameter
$\beta_{\rm g}$ in the torus as a maximum at $\beta_{\rm g,\max} :=
2p_{\max}/b_{\max}^2=100$, where $p_{\max}$ and $b_{\max}$ are the
maximum pressure and magnetic fields, respectively. Note that the
maximum of the total pressure and of the magnetic field do not coincide
and that, in order to excite the development of the MRI, the thermal
pressure of the torus is perturbed with a random noise with maximum
relative amplitude of $2\%$.

As customary in codes solving the equations of general-relativistic
hydrodynamics or of relativistic MHD \citep{Rezzolla_book:2013}, the
whole computational domain needs to be filled with a fluid, including the
regions outside the compact objects, which acts as a tenuous
atmosphere. In our simulations, the rest-mass density in the atmosphere
is initialised following two power laws, namely,
\begin{align}
\rho_{\mathrm{atm}} &=3 \times 10^{-4}\,(R/r)^{3/2}\,\rho_{\max}\,,&\\ 
p_{\mathrm{atm}} &=3 \times 10^{-6}\,(R/r)^{5/2}\,\rho_{\max}\,, &\\
v^i_{\mathrm{atm}} &=0\,.&
\end{align}
Furthermore, to prevent small fluctuations from developing in the
atmosphere, the rest-mass density, the pressure, and the velocity are
reset to their floor values whenever $\rho<1.001\,\rho_{\text{atm}}$
or $p<1.001\, p_{\text{atm}}$.

In order to increase the resolution in the radial direction, we make use
of a logarithmic radial coordinate, $s(r):=\ln r$, which is uniformly
spaced in the range $s\in [2.26,6.9]$, thus corresponding to a radial
coordinate $r\in[R,1000\,M]$. The angular dimension with $\theta \in
[0,\pi]$, on the other hand, is covered with a uniform grid, so that the
two-dimensional domain is covered at the coarsest level with an array of
$320\times 128$ grid cells. Furthermore, to further increase the
resolution efficiently, we employ a three-level adaptive mesh refinement
(AMR) based on a L\"ohner scheme \citep{Loehner87} to estimate the
errors. Additionally, after defining the MRI quality factor
as the ratio of the fastest growing MRI mode to the resolution in
the locally non-rotating reference frame (LNRF)
\begin{equation}
  Q^{(\theta)} := \frac{2\pi b^\mu \, \hat{e}_\mu^{(\theta)}}
  {\sqrt{\left(\rho h+b^2\right)}\left(u^\phi/u^t\right)}
  \frac{1}{\Delta \theta^\mu \, \hat{e}_\mu^{(\theta)}}\,,
  \quad \Delta\theta^\mu:=\left(0,0,\Delta \theta,0\right)\,,
\end{equation}   
where $\Delta\theta$ is the distance between two adjacent grids along the
$\theta$-direction and $\hat{e}_\mu^{(\alpha)}$ are the orthonormal bases
of the LNRF, we refine the grid wherever $Q^{(\theta)} < 5$
\citep{Sano2004, Noble2010, McKinney2012}.

Unlike black holes, whose event horizon acts as an absorbing null
surface, neutron stars have a physical timelike surface that needs to be
modeled suitably. A first possibility is to describe the neutron star as
a self-gravitating fluid and to describe the interaction of the accretion
flow onto the star as the interaction between two fluids. This approach
has a long history in numerical relativity \citep[see,
  \eg][]{Font02c,Baiotti04} but is not particularly convenient here for
at least two reasons. Firstly, it leads to very small timesteps in the
central cells of the star, which have a very small volume and a rather
simple dynamics. Secondly, the interaction of the accreting fluid with
the stellar surface and the large differences in their rest-mass
densities can lead to strong shocks in energy and velocity triggering
numerical artefacts.

A second possibility, and the one employed here, is to treat the stellar
surface as a surface across which suitable boundary conditions are
imposed. In particular, following \citet{Abarca2018} and
\citet{Takahashi2018}, at the inner boundary of the radial domain, we
impose a ``reflective'' boundary condition on the radial (and hence
normal) component of the fluid three-velocity, while no change is made to
the tangential components for conservation of linear momentum. More
specifically, we set
\begin{align}
&v^r_{\mathrm{out}} = -v^r_{\mathrm{in}}\,, &
&v^{\theta}_{\mathrm{out}} = v^{\theta}_{\mathrm{in}}\,, &
&v^{\phi}_{\mathrm{out}} = v^{\phi}_{\mathrm{in}}\,, &
\end{align}
where the subscripts ``in'' and ``out'' refer to the innermost cell in
the domain and to the ghost cells, respectively. On the other hand,
``continuous'' boundary conditions -- where the value of the innermost
cell is copied to all the ghost cells -- are applied to all components of
the magnetic field
\begin{equation}
B^i_{\mathrm{out}}=B^i_{\mathrm{in}}\left(s_{\min}\right)\,,
\end{equation}
and ``symmetric'' boundary conditions are used for all the other
primitive variables
\begin{equation}
\psi_{\mathrm{out}}=\psi_{\mathrm{in}}\,,
\end{equation}
where $\psi$ represents a generic scalar primitive variable. Furthermore,
reflective boundary conditions are also applied on the conservative
variables at the poles $\theta=0$ and $\theta=\pi$, \ie
\begin{align}
  &B^{r}_{\mathrm{out}} = B^{r}_{\mathrm{in}}\,,  &
  &S^{r}_{\mathrm{out}} = S^{r}_{\mathrm{in}}\,,  \\
  &B^{\theta}_{\mathrm{out}} = -B^{\theta}_{\mathrm{in}}\,, &
  &S^{\theta}_{\mathrm{out}} = -S^{\theta}_{\mathrm{in}}\,, \\
  &B^{\phi}_{\mathrm{out}} = B^{\phi}_{\mathrm{in}}\,,  &
  &S^{\phi}_{\mathrm{out}} = S^{\phi}_{\mathrm{in}}\,,  \\
  &\chi_{\mathrm{out}} = \chi_{\mathrm{in}}\,,  &
  &\phantom{S^{r}_{\mathrm{out}} = S^{r}_{\mathrm{in}}\,,}  &
\end{align}
where $S^i$ is the Eulerian momentum density and $\chi$ represents a
generic scalar conserved variable \citep{Porth2017,
  Olivares2019}. Finally, ``outgoing'' boundary conditions are applied to
the primitive variables at the outer radial boundary \ie
\begin{align}
&v^r_{\mathrm{out}} = \max\left(0,v^r_{\mathrm{in}}\left(s_{\max}\right)\right)\,, &
&v^{\theta}_{\mathrm{out}} = v^{\theta}_{\mathrm{in}}\left(s_{\max}\right)\,, &
&v^{\phi}_{\mathrm{out}} = v^{\phi}_{\mathrm{in}}\left(s_{\max}\right)\,,  \\
&B^r_{\mathrm{out}} = B^r_{\mathrm{in}}\left(s_{\max}\right)\,, &
&B^{\theta}_{\mathrm{out}} = B^{\theta}_{\mathrm{in}}\left(s_{\max}\right)\,, &
&B^{\phi}_{\mathrm{out}} = B^{\phi}_{\mathrm{in}}\left(s_{\max}\right)\,,  \\
&\psi_{\mathrm{out}} = \psi_{\mathrm{in}}\left(s_{\max}\right)\,, &
&\phantom{B^{\theta}_{\mathrm{out}} = B^{\theta}_{\mathrm{in}}\left(s_{\max}\right)\,,} &
&\phantom{B^{\phi}_{\mathrm{out}} }  
\end{align}

As a concluding remark, we note that all of the simulations performed
have been evolved up to a time $t=75000\,M_{\odot}\simeq
370\,\mathrm{ms}\simeq 3.39\,\tau_0$ where $\tau_0$ is the initial
orbital period at $r_{\mathrm{max}}$.

\begin{figure*}
\includegraphics[width=0.85\textwidth]{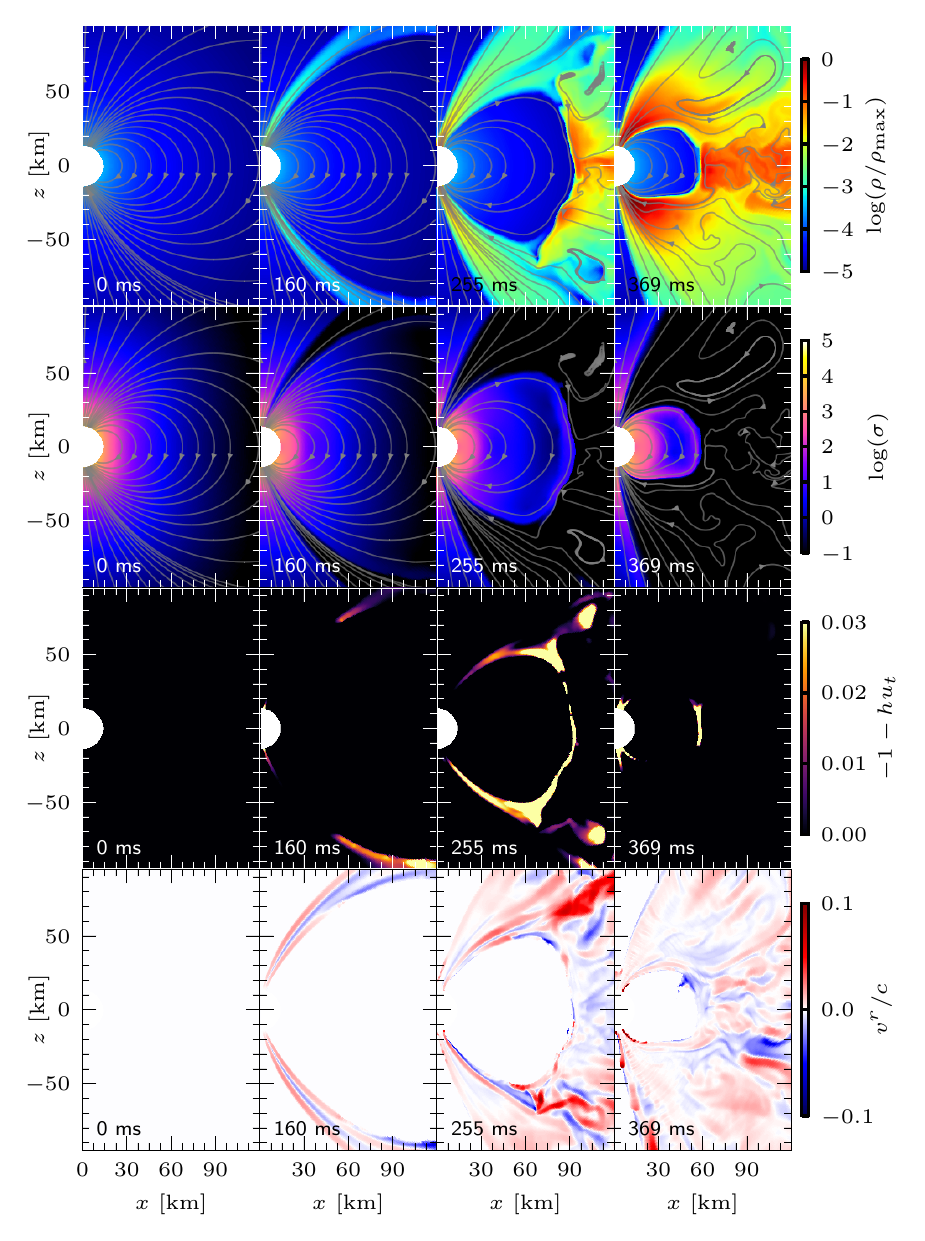}
\caption{Rest-mass density, magnetisation, Bernoulli parameter, and
  radial velocity at various times in the simulation with $B_9=5$. The
  grey lines denote the magnetic field lines. \label{fig:ss_zoom}}
\end{figure*}

\begin{figure*}
\includegraphics[width=0.85\textwidth]{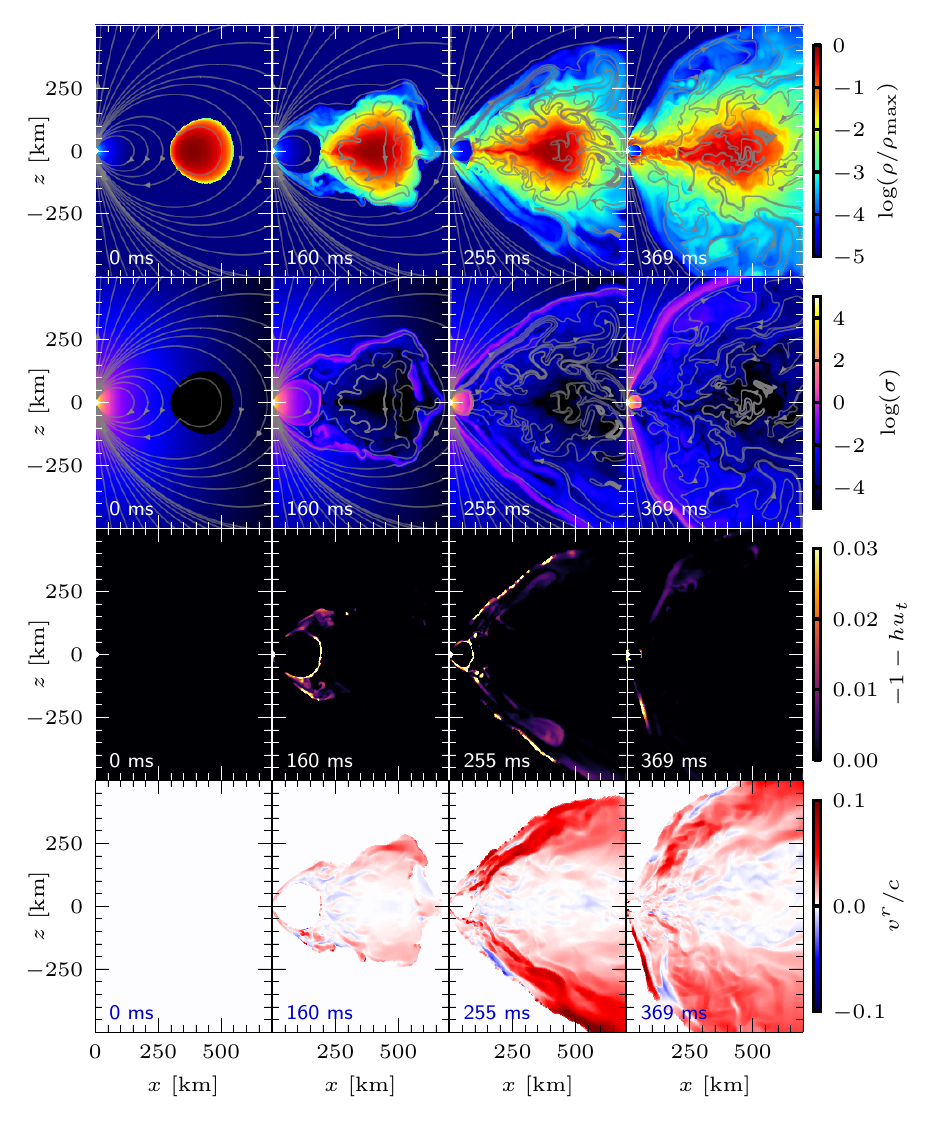}
\caption{Same as Fig.~\ref{fig:ss_zoom} but over a larger domain (still
  smaller than the computational one) to highlight the large-scale
  dynamics of the plasma. }
\label{fig:ss_large}
\end{figure*}

\section{Results}
\label{sec:Re}

In what follows we discuss in detail the various aspects of our
simulations, concentrating first on the plasma dynamics, so as to move,
later on, on the properties of the angular velocity in the accreting flow
and of the pitch factor across the torus.

\subsection{General plasma dynamics}

Driven by MRI, the matter in the torus flows towards the star advecting
with it the magnetic field lines. From time to time, and responding to
the turbulent nature of the accretion process, the inflow is halted by
the concentration of magnetic field lines that accumulate around the
magnetospheric radius. However, as additional matter continues to inflow
and accumulates, the magnetic-pressure barrier is overcome, restoring the
accretion onto the star. This final part of the accretion process --
namely the one made by the plasma before reaching the stellar surface --
does not proceed along the equatorial plane, but along the magnetic tubes
produced by the most external magnetic field lines of the stellar
magnetosphere (see Figs. \ref{fig:ss_zoom} and \ref{fig:ss_large}), which
become considerably stronger near the star and thus channel the matter
from the accretion torus onto the stellar poles.

This process of mass transfer from the torus to the stellar surface
starts well before the bulk of the torus has reached the magnetospheric
radius and a quasi-stationary configuration is reached. Indeed, with the
exception of the case with a weak stellar magnetic field, \ie for
$B_9=1$, the matter of the torus reach the stellar surface only along the
flux tubes in the polar regions and not with a direct accretion along the
equatorial plane. As a representative example, snapshots from the
simulation with $B_9=5$ are presented in Figs.~\ref{fig:ss_zoom} and
\ref{fig:ss_large}, whose different columns refer to snapshots at
different times during the evolution, while the different rows report the
rest-mass density (first row from the top), the magnetisation (second
row), the Bernoulli parameter (third row) and the radial component of the
three velocity (fourth row).

\begin{figure*}
\center
\includegraphics[width=0.95\columnwidth]{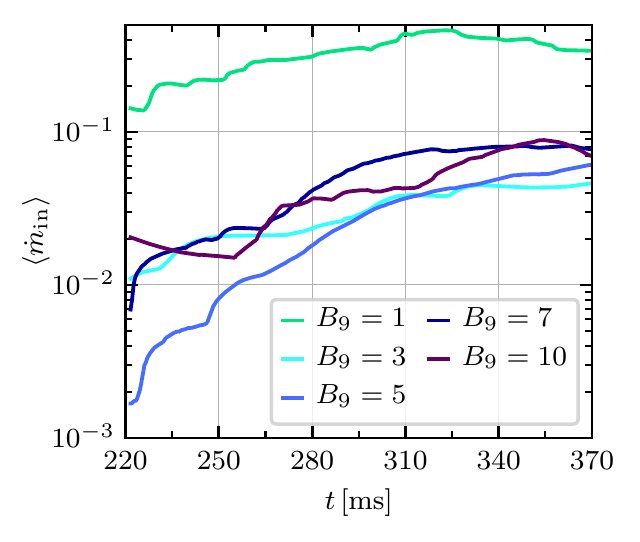}
\hskip 1.0cm
\includegraphics[width=0.95\columnwidth]{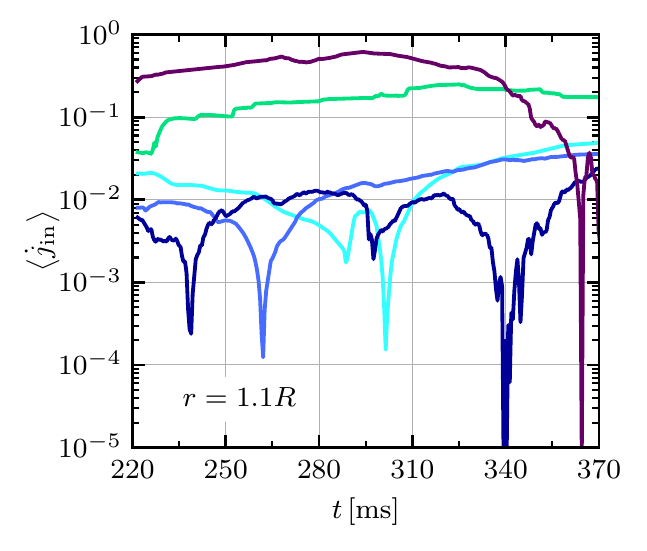}
\caption{Mass-accretion rate (left panel) and angular-momentum transport
  rate (right panel) for all simulations calculated at $r=1.1\,R$. They
  are presented in dimensionless units as given in Eq.~\ref{eq:units} and
  their running averages are taken in $50\,\mathrm{ms}$ window.
\label{fig:MJdot}}
\end{figure*}

Together with the inflow, the dynamics of the torus is accompanied by a
net outflow of matter, some of which is actually gravitationally unbound
and launched at large distances. The ejection of matter can take place
for at least three different reasons. First, as mentioned above, a
significant amount of magnetic reconnection takes place at the outer
parts of the stellar magnetosphere and around the equatorial plane. The
consequent conversion of magnetic energy into internal energy can be so
efficient (see third rows in Figs.~\ref{fig:ss_zoom} and
\ref{fig:ss_large}) such that the matter in the torus becomes
gravitationally unbound. Second, matter becomes heated as it is
channelled in the tight flux tubes on the stellar poles, again reaching
internal energies that can, episodically, lead to matter
ejection. Finally, some of the accreting matter is simply reflected from
the stellar surface, especially when the density in the polar flux tubes
decreases because of a smaller accretion rate and the matter can freely
fall onto the star.  Under these conditions, strong shocks can develop at
the stellar surface triggering powerful outflows with velocity
$>0.1$. Clearly, given the nature of these processes, which are basically
triggered by stochastic magnetic reconnection, the ejection of matter is
rather episodic and it is the most copious in the time window $\sim
200-300\,\mathrm{ms}$, that is, in the transition between the initial
inflow and the reaching of a stationary solution. Overall, we do not find
a periodic or quasi-periodic behaviour in the mass-ejection process in
our simulations.

A couple of remarks are useful at this point. First, we observe that some
of the matter escapes from the flux tubes as a result of magnetic
reconnection and forms magnetic islands (plasmoids). The
presence of these plasmoids can be appreciated, in particular, in the
third-column, third-row panels of Figs.~\ref{fig:ss_zoom} and
\ref{fig:ss_large}, where these magnetic islands are particularly
visible. Although the magnetisation of these magnetic islands is
larger than that of the rest of plasma, it is nevertheless modest and of
order unity. Nevertheless, these structures share many of the properties
of the plasmoids found in accreting supermassive black holes
\citep{Nathanail2020, Nathanail2021b}. Second, because the stellar
magnetic field around the torus is weaker than the one initially seeded
in the torus, the latter expands due to magnetic buoyancy as the
simulation proceeds, giving rise to a low-density magnetised plasma, \ie
a ``corona'' (see top rows of Figs.~\ref{fig:ss_zoom} and
\ref{fig:ss_large}). The plasma in this corona can then be energised by
the matter reflected off the stellar surface and thus become unbound,
leaving the system (see bottom rows of Figs.~\ref{fig:ss_zoom} and
\ref{fig:ss_large}).

\begin{figure*}
\center
\includegraphics[width=0.95\columnwidth]{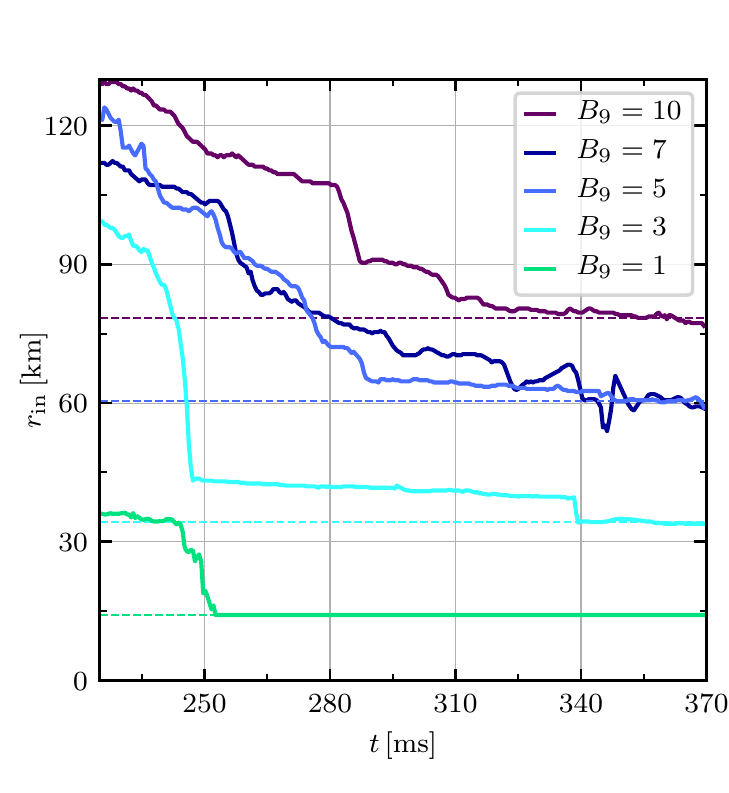}
\hskip 1.0cm
\includegraphics[width=0.95\columnwidth]{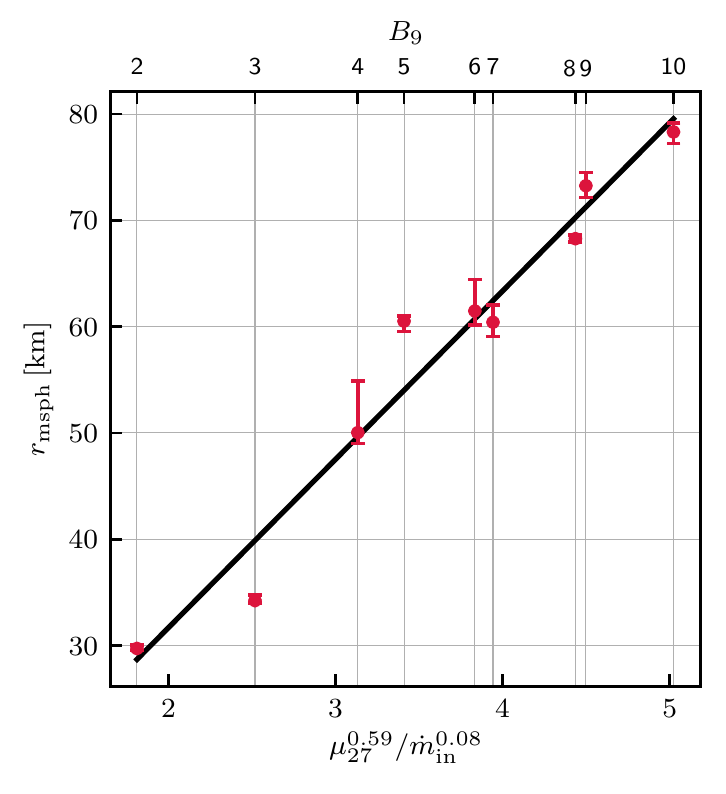}
\caption{Left panel: Evolution of the magnetospheric radius of the torus
  in time. The horizontal dashed lines denote the inner radius of the
  torus in steady state. Right panel: The magnetospheric radius vs.\ the
  best-fitting estimate of the magnetic dipole moment over the mass
  accretion rate. The top axis corresponds the magnetic field strength
  of the each simulation. The red points are the average of the last
  $20$ ms of the inner radius and error bars denote the maximum and
  minimum values in last $20$ ms. The black line is the best-fitting of
  the magnetic radius. The result for the $B_9=1$ simulation is excluded
  from the fitting since the torus extends to the star in this case.
\label{fig:ralven}}
\end{figure*}

To determine whether a fluid element is unbound we use the so-called
``Bernoulli parameter'', which we choose to set to $\mathrm{Be}:=-(1+h
u_t)$, so that fluid elements with a positive Bernoulli parameter can
escape to infinity\footnote{We note that several definitions
are possible for the Bernoulli parameter \citep[see][for a discussion
of different measurements of the outflowing material in a numerical
simulation]{Bovard2016} and that numerical simulations of accretion
flows either neglect inertial terms using a purely kinematic criterion
\citep[see, \eg][]{Porth2019}, or add contributions from the magnetic
field \citep[see, \eg][]{Narayan2012}. Because both approaches tend to
normally increases the amount of ejected matter, we here take a rather
conservative view and thus employ a definition that takes into account
inertial terms but only at the hydrodynamical level}. Given the
duality in the inflowing and outflowing material, we measure the
``inflowing'' mass-accretion rate in terms of the matter that is moving
radially inwards, namely as
\begin{equation}
\dot{M}_{\mathrm{in}} := - 2\pi\int \rho u^r
\sqrt{-g}\,d\theta\,,
\label{eq:mdot}
\end{equation}
where the integrand refers to matter with $u^r<0$. Using similar
filtering criteria, the rate of inflowing angular momentum is defined as
\begin{align}
&\dot{J}_{\mathrm{in}} := -2\pi\int T^r_{\,\,\phi}
\sqrt{-g}\,d\theta \,\,&
\notag \\
&\phantom{\dot{J}_{\mathrm{in}}}\,
= -2\pi\int \left[\left(\rho h + b^2\right)u^r
  u_\phi -b^r b_\phi\right]
\sqrt{-g}\,d\theta\,,& 
\label{eq:jdot}
\\
&\dot{J}_{\mathrm{in,matt}} := -2\pi \int \rho h u^r u_\phi
\sqrt{-g}\,d\theta\,,&
\label{eq:jdotmat} 
\end{align}
where $T^r_{\,\,\phi}$ is the only relevant component of the
energy-momentum tensor and $\dot{J}_{\mathrm{in,matt}}$ refers to the
portion of the angular-momentum transfer rate related to the matter
only.

When inspecting the evolution of the mass-accretion rate at a high
cadence we observe that it exhibits large fluctuations around an average,
but with secularly varying value. These fluctuations reflect the
turbulent and chaotic nature of the accretion flow, which is triggered
both by the development of the MRI and by the reconnection processes
taking place at the edge of the stellar magnetosphere.

Figure \ref{fig:MJdot} reports the evolution of the inflowing
mass-accretion rate (left panel) and of the angular-momentum transfer
rate (right panel), when expressed in terms of normalised quantities [see
  Eqs. \eqref{eq:units}], and for some representative values of the
stellar magnetic field\footnote{We recall we have considered ten
different magnetisations, \ie $B_9=1-10$, but report only five in
Fig. \ref{fig:MJdot}.} Both rates are measured across a spherical surface
with coordinate radius $r=1.1\,R$ and we have removed the smallest
fluctuations by performing a running average over a time window of
$\pm 50\,\mathrm{ms}$.

Overall, two main effects emerge. First, and as it is natural to expect,
the mass-accretion rate grows from its very small initial values and
settles to an almost constant rate as a quasi-stationary accretion
process is reached. There are two factors that make the $B_9=1$ case
different from the others. The first is that the weaker stellar magnetic
field offers a smaller resistance to the accretion flow, which can reach
the stellar surface more easily, thus increasing the mass-accretion
rate. The second and possibly more important factor is that that
accretion in the $B_9=1$ is not mediated by closed magnetic field lines
and indeed matter from the torus can reach the stellar surface already
around the equatorial plane, thus increasing the efficiency of the
accretion process.

The transport of angular momentum, on the other hand, is far less
regular, occasionally switching to negative values, and there is
considerable variability associated both with the specific phase of the
accretion process and with the stellar magnetic field. Neutron stars with
larger magnetic fields will collect matter from regions that are farther
out in the torus -- the magnetospheric radii are larger for larger
stellar magnetic fields -- and since the specific angular momentum grows
with the radius ($\ell \propto \Omega r^2 \sim r^{1/2}$ for Keplerian
flows with $\Omega \propto r^{-3/2}$), the accreting flow will transfer
larger amounts of angular momentum. On the other hand, larger magnetic
fields will also produce more violent episodes of reconnection at the
magnetospheric radius, possibly preventing accretion and thus the
transfer of angular momentum, and explaining the variability of the
angular-momentum transfer rate recorded in the case of $B_9=10$ (see
right panel of Fig. \ref{fig:MJdot}. Finally, we note that for large
stellar magnetic field the ``magnetic component'' of the angular-momentum
transfer rate, \ie that proportional to $b^2u^ru_\phi-b^r b_\phi$ in
Eq.~\eqref{eq:jdot}, dominates over the ``matter part'', \ie the one
proportional to $\rho h u^ru_\phi$ in Eq.~\eqref{eq:jdot}. These magnetic
components are clearly very sensitive to the reconnection taking place at
the edge of the magnetosphere, but are only weakly correlated to the
variations of the mass-accretion rate. In our setup, the threshold
magnetic field for this transition appears to be given by the
configuration with $B_9\simeq5$. Configurations with smaller magnetic
fields have smoother magnetospheres and experience comparatively smaller
(negative) contributions from the magnetic component of the
angular-momentum transfer rate; hence, they transfer larger amounts of
angular momentum overall (see right panel of Fig.~\ref{fig:MJdot}).

The episodic nature of the reconnection processes that trigger both
inflow and outflow can be best appreciated when considering the tight
(nonlinear) correlation between the mass-accretion rate and the accretion
rate of angular momentum. The existence of a tight connection between the
stellar spin and the mass-accretion rate is expected because a number of
a observations \citep[see, \eg][]{Bildsten97, Doroshenko2018, Zhang2019e,
  Ji2020} clearly show a correlation between the spin-up rate of the star
and the luminosity. In our simulations we do not compute the X-ray
luminosity, nor we measure the spin up of the star, which is always
treated as nonrotating. However, we do measure both the mass-accretion
rate -- which is naturally associated to the X-ray luminosity, the latter
being larger for more copious mass-accretion rates -- and the accretion
of angular momentum, which can naturally be associated with changes in
the spin of the star. Indeed, using the angular-momentum transfer rate
measured from the simulations, we estimate that the changes induced in
the stellar spin would be of the order of $\dot{\Omega} \simeq
10^{-12}\,\mathrm{Hz\,s^{-1}}$ for $B_9=10$, which is in very good
agreement with the spin-up rates measured in observations, \ie
$\dot{\Omega}\simeq 10^{-12}-10^{-11}\,\mathrm{Hz\,s^{-1}}$
\citep{Sugizaki2017}. Hence, it is reasonable to assume a correlation of
the type
\begin{equation}
  \label{eq:jdot_mdot}
  \dot{j}_{\mathrm{matt, in}}\propto\dot{m}_{\mathrm{in}}^{\lambda}\,,
\end{equation}
where the exponent $\lambda$ has been estimated in various analytical
models and, for instance, is set to $\lambda=0.86$ in the Ghosh \& Lamb
model \citep{Ghosh1979}, $\lambda=0.9$ in \citet{Kluzniak2007} and
$\lambda=0.64$ in \citet{Shakura2012}. We note that we here
consider only the matter part of the angular-momentum flux, thus
neglecting the magnetic contributions to the total torque. The latter,
in fact, are produced at the edge of the magnetosphere and, being
mostly stochastic in nature, are not correlated with the mass-accretion
rate and thus cannot be modelled with the simple ansatz
given in Eq.~\eqref{eq:jdot_mdot}.

\begin{figure}
\includegraphics[width=0.95\columnwidth]{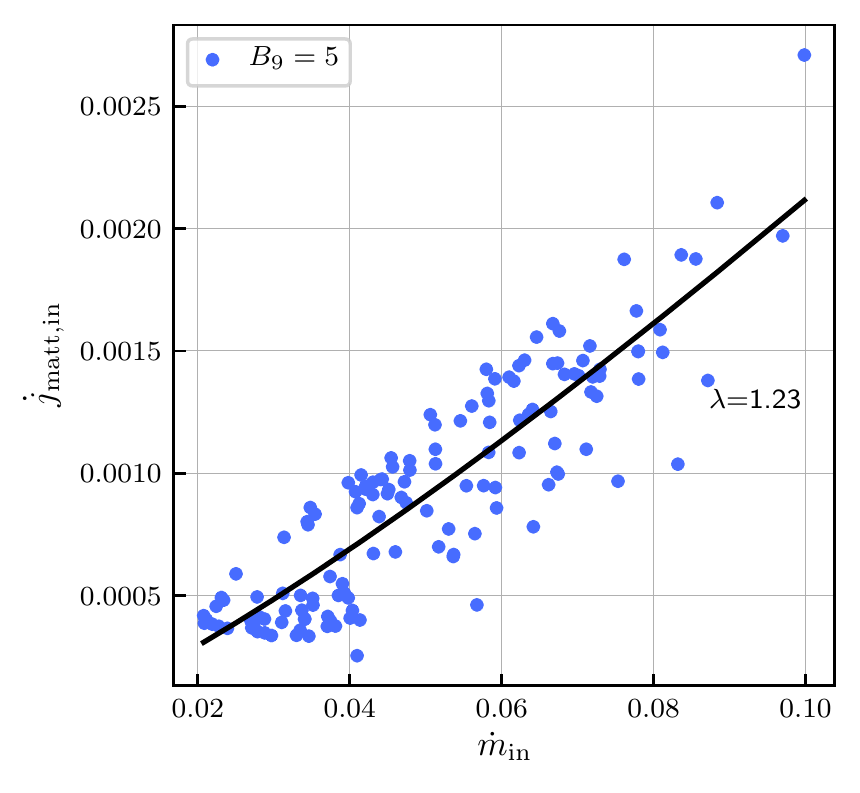}
\caption{The matter part of the angular momentum transport rate
  $\dot{j}_{\mathrm{matt, in}}$ [see Eq. \eqref{eq:jdot_mdot}] shown
  as function of the mass-accretion rate for the $B_9=5$ simulation.}
 \label{fig:jmdot_correlation}
\end{figure}

Figure \ref{fig:jmdot_correlation} reports the behaviour of the
mass-accretion and of the angular-momentum transfer rates for the
representative case with stellar magnetic field $B_9=5$ and relative to
the quasi-stationary part of the evolution, \ie for $t >
300\,\mathrm{ms}$; very similar correlations are found in all cases
considered. As suggested by Eq. \eqref{eq:jdot_mdot},
Fig. \ref{fig:jmdot_correlation} shows a clear nonlinear correlation
between these two quantities, highlighting that not only the steady-state
accretion of matter should lead to a steady a variation of the stellar
spin, but also that fluctuations in $\dot{j}_{\mathrm{matt, in}}$ are
directly related to fluctuations in $\dot{m}_{\mathrm{in}}$. This
behaviour provides strong support to the idea that the episodic
reconnection processes taking place at the edge of the magnetosphere
should lead both to an increased luminosity and to a stellar spin-up.

Using the various simulations performed, we have estimated the values of
the correlation exponent $\lambda$, after removing from our data-sets
those variations\footnote{We note that there is no excluded
  data for our representative example $B_9=5$ shown in
  Fig.~\ref{fig:jmdot_correlation}.} in the mass-accretion rate or in
the angular-momentum transfer rate that are larger by a factor of seven
with respect to the corresponding average values. In this way we can
filter out the most extreme fluctuations and find that the exponent
varies in the range $0.93 \lesssim \lambda \lesssim 1.43$, with a average
value of $\langle \lambda \rangle = 1.18$ and no clear correlation with
the value of the stellar magnetic field. Interestingly, this result is in
very good agreement with the analysis of $12$ X-ray sources reported by
\citep{Sugizaki2017}, where $\lambda$ is estimated to be $1.03$.

\subsection{Inner edge of the torus and magnetospheric radius}
\label{sec:rmsph}

A very important quantity in our analysis is the determination of the
magnetospheric radius $r_{\rm msph}$, namely of the radius at which the
disc accretion on the equatorial plane, is diverted into the polar flux
tubes reaching the stellar surface. Clearly, the location of the inner
radius $r_{\rm in}$ where this process takes place will depend not only
on the strength of the stellar magnetic field, but also on whether a
quasi-stationary equilibrium between the accretion flow and the matter
ejection is reached. Hence, we measure the magnetospheric radius as the
asymptotic value attained by the inner radius when a quasi-stationary
flow has been established. Stated differently, we define the
magnetospheric radius as $r_{\rm msph} = \lim_{t \to t\gg \tau_0} r_{\rm in}$
and monitor the evolution of the inner radius till reaches a stationary
value.

\begin{figure}
\includegraphics[width=1.0\columnwidth]{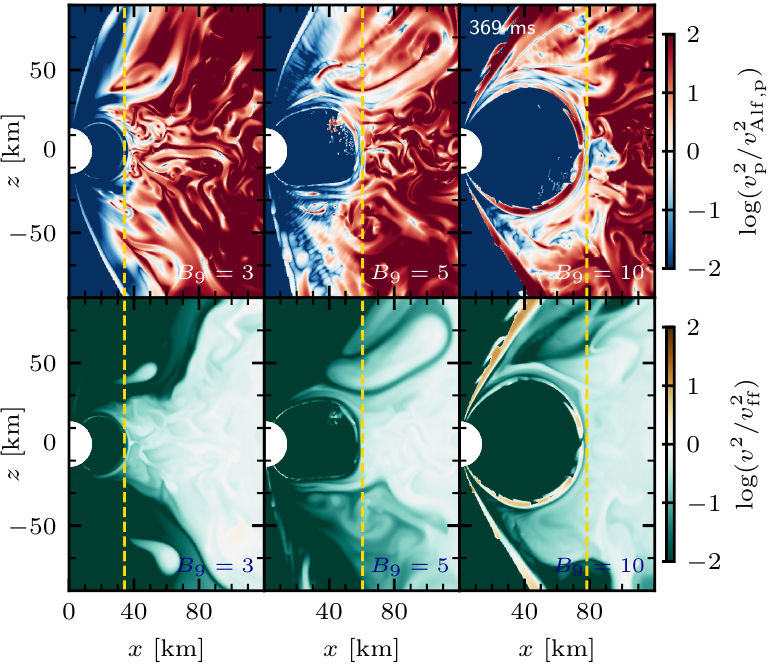}
\caption{The ratio of the square of 
the poloidal velocity $v_{\mathrm{p}}^2:=v^r
v_r+v^\theta v_\theta$ to the poloidal Alfv\'{e}n
velocity [Eq.~\eqref{eq_valf}] (top row), and the
Eulerian velocity $v^2=v^i v_i$ to the free-fall velocity
[Eq.~\eqref{eq_vff}] (bottom row) at the final time of three
representative simulations. The dashed vertical lines mark the
magnetospheric radii in each simulation.}
\label{fig:velocity} 
\end{figure}

There are several different ways in which the inner radius can be
extracted from the numerical simulations and this is because the edge of
the magnetosphere is always very straightforward to determine when
looking either at the rest-mass density (see Figs.~\ref{fig:ss_zoom} and
\ref{fig:ss_large}) or at the velocity field (see
Fig.~\ref{fig:velocity}). Out of these possibilities -- all of which
yield very similar results -- we adopt the simplest one, namely, we set
the inner radius to be the location where, moving along the equatorial
plane out from the stellar surface, the rest-mass density reaches the
threshold value $\rho_{\rm th}=10^{-2}\,\rho_{\max}$. This radius
obviously changes with time and we report in the left panel of
Fig.~\ref{fig:ralven} its evolution for the same representative values of
the magnetic field shown in Fig.~\ref{fig:MJdot}. Using the values of
$r_{\rm in}$, we therefore calculate the magnetospheric radius by taking
the corresponding value when averaged over the last $20\,{\mathrm{ms}}$
of the simulation (see left panel of Fig.~\ref{fig:ralven}).

The importance of the magnetospheric radius is that its value can be
estimated already in Newtonian gravity and is given by 
\begin{equation}
  r_{\mathrm{msph}} :=\xi \left(\frac{\mu^4}{8M\dot{m}^2}\right)^{1/7}\,,
\label{eq:rmsph}
\end{equation}
where $\mu$ is the magnetic moment of a star mass with mass $M$ and
subject to a mass-accretion rate $\dot{m}$. In expression
\eqref{eq:rmsph}, $\xi$ is a dimensionless coefficient of order unity
whose value depends on the assumptions about the details of the
disc-magnetosphere interaction, \ie the width of the zone where the
magnetic field of the star can penetrate the disc and the physical
processes limiting the growth of the toroidal field generated by the
stellar field lines penetrating the disc. For instance, $\xi = 0.5$ in
the model of Ghosh \& Lamb \citep{Ghosh1979a}, while it is estimated to
be $0.3-1.2$ in other studies \citep{Wang1987, Wang1996, Psaltis1999b,
  Erkut2004, DallOsso2016}. Moreover, $\xi$ might depend on the ratio
$r_{\mathrm{msph}}/r_{\mathrm{co}}$, where $r_{\mathrm{co}} :=
(M/\Omega_*^2)^{1/3}$ is the corotation radius, with $\Omega_*$ the
angular velocity of the star, and on the inclination angle between
rotation and magnetic axis \citep{Bozzo2018}.

In order to verify whether expression \eqref{eq:rmsph} is valid also in a
non-trivial and general-relativistic context, we have a more generic
expression of the magnetospheric radius, which includes information on
the mass-accretion rate, via a power-law of the type 
\begin{equation}
  \label{eq:rmsph_gr}
   r_{\rm msph}= a_1\,
   \left(\mu_{27}\right)^{a_2}\,\left(\dot{m}_{\mathrm{in}}\right)^{a_3}
   \left(\frac{M}{M_{\odot}}\right)\,,
\end{equation}
where the inflow mass-accretion rate $\dot{m}_{\mathrm{in}}$ has been
measured at $r=1.1\,R$ and time-averaged over the last
$110\,{\mathrm{ms}}$. We have reported in the right panel of
Fig.~\ref{fig:ralven} the measured values of the magnetospheric radius
and estimated the fitting coefficients via the ansatz
\eqref{eq:rmsph_gr}. In this way, we have found that
\begin{equation}
\label{eq:rmsph_gr_coeffs}
  a_1=7.66\pm 1.6\,, \qquad a_2=0.59\pm 0.06\,, \qquad a_3=-0.08\pm 0.06\,.
\end{equation}
Interestingly, while the magnetospheric radius depends on the strength of
the stellar magnetic field with the almost same power-law as in the
Newtonian expression (the latter predicts that $a_2=4/7\simeq 0.571$), we
find that the correlation between the mass-accretion rate and the
location of the magnetospheric radius is a weaker than in the Newtonian
estimate (the latter predicts that $a_3=-2/7\simeq -0.286$). More
importantly, however, the relevance of expressions \eqref{eq:rmsph_gr}
and \eqref{eq:rmsph_gr_coeffs} is that they do not depend on the
idealised, stationary, axisymmetric assumptions behind the Newtonian
expression \eqref{eq:rmsph}. Rather, they reflect the quasi-stationary
accretion flow that may be present near an accreting pulsar, where all
quantities have a fully turbulent nature and the magnetic field does not
have a simple power-law scaling.

In Appendix~\ref{sec:MagRad} we will provide an analytic expression for
the magnetospheric radius for a spherically symmetric accretion flow onto
a magnetised star up to the second order in the relativistic corrections
$\mathcal{O}(M/r)$. Also in this case we find that $r_{\rm msph} \propto
\mu^{4/7}$, but also that the magnetospheric radius is smaller than the
Newtonian counterpart for the same set of parameters [see
Eq. \eqref{eq:rmsph_gr_appendix}]. This is indeed rather natural as the
general-relativistic gravity will be more intense and require a
comparatively larger magnetic field to push out the magnetospheric
radius.

\subsection{Angular velocities in the torus}

Although the flow inside the torus is nonrelativistic, matter in the
outer parts of the corona can reach quite high speeds, \ie ($\gtrsim
0.1\,c$ (see Figs.~\ref{fig:ss_zoom} and \ref{fig:ss_large}). A
convenient way to present the velocity field is to compare it with two
other characteristic velocity scales, namely, the poloidal
Alfv\'en velocity
\begin{equation}
  \label{eq_valf}
  v_{\mathrm{Alf,p}} :=\sqrt{\frac{B^r B_r + B^\theta B_\theta}{\rho}}\,,
\end{equation}
and the free-fall velocity in a spherically symmetric metric of mass $M$
\citep{Rezzolla_book:2013}
\begin{equation}
  \label{eq_vff}
  v_{\mathrm{ff}}:=\sqrt{\frac{2M}{r}\left(1-\frac{2M}{r}\right)}\,.
\end{equation}

Figure \ref{fig:velocity} provides a view of the velocity
field in terms of the square of the poloidal velocity,
$v_{\mathrm{p}}^2:=v^r v_r+v^\theta v_\theta$, normalised to the
poloidal Alfv\'en velocity (top row) and the Eulerian velocity,
$v^2:=v^i v_i$, normalised to the free-fall velocity (bottom row).
Note that the flow is super-Alfv\'enic (and supersonic) up to the
interaction between the accretion flow and the magnetosphere, and it
becomes sub-Alfv\'enic (and subsonic) inside the funnel, where however
the matter content is very small. Similarly, matter in the torus and near
the equatorial plane moves towards the stellar surface with speeds larger
than the free-fall velocity. On the other hand, a significant
acceleration is experienced by the plasma as it reaches the stellar
surface along the funnel walls.

\begin{figure}
\includegraphics[width=1.0\columnwidth]{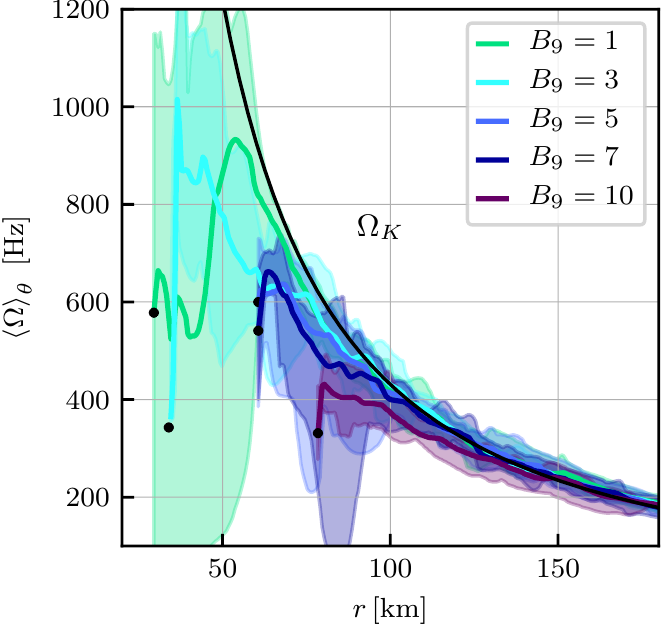}
\caption{Radial profiles of the $\theta$-averages of the angular velocity
  for some representative stellar magnetic fields (solid lines of
  different colours). All lines start at the magnetospheric radius and
  the corresponding shaded regions mark the maximum and minimum values of
  the angular velocity at a given radius. Shown with a black solid line
  is the profile of the Keplerian angular velocity, indicating that the
  flow is sub-Keplerian near the magnetosphere and essentially Keplerian
  at large distances. Finally, filled black circles mark the position of
  the various magnetospheric radii.}
\label{fig:Omega}
\end{figure}

Also interesting is to consider the radial profiles of the angular
velocity in the torus. While the motion there is very turbulent, it is
nevertheless possible to perform a polar average defined as 
\begin{equation}
  \left\langle\Omega\right\rangle_\theta(r) :=
  \frac{\int \Omega (r,\theta)\sqrt{-g}\,d\theta}{\int \sqrt{-g}\,d\theta}\,,
  \label{eq:Omega}
\end{equation}
where, in order to exclude any contribution from the corona, we have
restricted the integrals to a wedge of $\sim 20\degr$ around
the equatorial plane and to densities in the range
$\rho/\rho_{\mathrm{max}}>10^{-2}$. The corresponding angular velocities
at the final moments of the simulation are presented in
Fig.~\ref{fig:Omega} for some of the representative stellar
magnetic-field strengths. Note that, in all cases, the averaged angular
velocity profiles have a global maximum near to the inner edge of the
torus.

Following \citet{Ghosh1977}, we define the \textit{transition region}
as the region between the magnetospheric radius and the location where the
stellar dipolar magnetic field modifies the angular velocity away from a
Keplerian profile. Using this definition across the various cases of
stellar magnetic fields, we find that the width of the transition region
varies between $2$ and $9\,{\rm km}$. More importantly, we find that the
width of the transition region is not correlated with the strength of the
stellar magnetic field. Not surprisingly, given the turbulent nature of
the accretion flow and the sharp transition between the torus and the
corona, the angular velocity can become locally negative (see
Fig.~\ref{fig:Omega}). However, this behaviour is not stationary and over
a time average the $\theta$-averaged angular velocity in Eq.~\ref{eq:Omega}
 is always positive at all latitudes considered.

Although no reflection symmetry is imposed in the simulations and the
accretion flow is highly turbulent, matter is channelled onto almost
symmetrical locations at the north and south hemispheres of the star, \ie
at latitudes of $15\degr-31\degr$ and $163\degr-152\degr$,
respectively. The exact position of these ``hot spots'' obviously depends
on the strength of the stellar dipolar magnetic field, so that the
elevation increases with the stellar magnetic field; this is to be
expected when considering the Newtonian estimate of the boundary of the
polar cap $\theta_{\rm c}$, \ie $\sin \theta_{\rm c} = \sqrt{R/r_{\rm
    msph}}$; note that the Newtonian estimate for $\theta_{\rm c}$ is
$25-45\%$ larger than what we measure in our general-relativistic
calculations. At the same time, the fraction of the stellar surface where
matter can accrete becomes progressively smaller as the stellar magnetic
field increases. More specifically, the fraction of the stellar surface
area interested by the accreting plasma can be large as $28\%$ for the
$B_9=2$ models, while it reduces to only $5\%$ when considering the
$B_9=10$ model. 

\begin{figure*}
\includegraphics[width=0.9\textwidth]{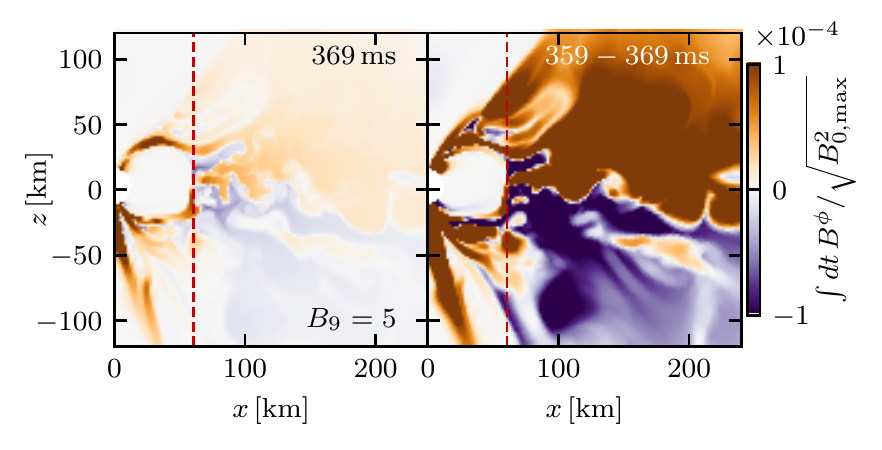}
\caption{The snapshot of the toroidal magnetic field for the $B_9=5$ case
  at $369\,\mathrm{ms}$ (left panel) and its time integration over the
  last $10\,\mathrm{ms}$ (right panel). The red dashed lines stand for
  the magnetospheric radius.}
    \label{fig:bphi}
\end{figure*}

\subsection{Secular toroidal magnetic field and pitch factor}

We have already discussed how the development of the MRI is responsible
for the transport of angular momentum and the accretion of matter towards
the stellar magnetosphere. Also, while the initial magnetic field is
purely poloidal, the turbulent motion produced by the MRI generates, as a
result of the large conductivity of the plasma, a toroidal magnetic field
on those lengthscales over which the instability develops. This is a
well-known process \citep{BalbusHawley1998}, which has been studied in
great detail in the accretion onto supermassive black holes \citep[see,
  \eg][]{Porth2019_etal}. However, accretion onto a magnetised star
introduces an important difference with respect to the accretion onto a
black hole, and this is in the generation of a globally coherent, secular
toroidal magnetic field in the disc. The source of this difference is
indeed the presence of a stellar magnetic field, which is clearly absent
in the case of a neutrally charged hole, and that add to the locally
turbulent toroidal magnetic field and is stretched in the azimuthal
direction by the global rotation of the accretion disc \citep[see][for an
analogous process in the case of a $r$-mode unstable neutrons
star]{Rezzolla00}.

The growth of this magnetic field can be easily deduced from the
Newtonian expression of the induction equation for the toroidal magnetic
field that, near the equatorial plane, where the radial component of the
magnetic field is much smaller than the polar one, \ie $B^{r} \ll
B^{\theta}$, reads
\begin{equation}
  \label{eq:Bphi}
\left\langle{\partial_t B^\phi}\right\rangle_t = \frac{1}{r}
\left\langle\partial_\theta\left(v^\phi B^\theta\right)\right\rangle_t \,,
\end{equation}
where the $\langle\ \rangle_t$ brackets indicate a time average. Hence,
toroidal magnetic field is produced by the shearing of the poloidal
magnetic field, grows linearly in time, at least in the ideal-MHD
approximation of infinite conductivity, and changes polarity across the
equator.

In the magnetically threaded disc model
\citep{Ghosh1979a,Wang1987,Campbell1992}, the toroidal magnetic field is
generated as a result of the angular velocity difference between the
star, $\Omega_*$, and the rotating plasma
\footnote{The mentioned models give the steady-state toroidal magnetic
field by the balancing of the advection and diffusivity terms in the
induction equation. Yet, we solve the GRMHD equations within the
ideal-MHD approximation, therefore, there is no mechanism to limit the
growth of the toroidal magnetic field in our numerical setup.}, hence,
the field is generated at a rate,
\begin{equation}
\left\langle \partial_t B^\phi \right\rangle_t = 
\left\langle \frac{1}{r}\left(\frac{\partial
  v^\phi}{\partial\theta} B^\theta\right)\right\rangle_t
\propto \pm \left\langle \left(\Omega-\Omega_*\right) B^\theta   \right\rangle_t
   \,,
\end{equation}
where ``$\pm$'' arises due to the vertical gradient switching sign
above and below the disc midplane. 

The generated toroidal magnetic field is illustrated in
Fig.~\ref{fig:bphi}, which reports a snapshot of the toroidal magnetic
field at time $t=369\,{\rm ms}$ (left panel) normalised to the initial
stellar value, but also its time integration over the last $10\,{\rm ms}$
(right panel). Clearly, the instantaneous toroidal magnetic field changes
polarity also on small scales and does not show a globally coherent
structure. This behaviour is very similar to that produced in simulations
of accretion discs onto black holes \citep[see, \eg][]{Nathanail2020} and
onto magnetised neutron stars \citep{Naso2011}. However, when averaged
over sufficiently long timescales, it is possible to appreciate also the
appearance of a clear polarity change across the equatorial
plane.

Another useful diagnostic of the properties of the accreting plasma is
the so-called ``pitch factor'', namely, the polar-averaged ratio of the
toroidal magnetic field to the poloidal magnetic field. Here we compute
this quantity at $r = 81\,\mathrm{km}$\footnote{This radial position has
been chosen as the one where the fluctuations in the pitch factor are the
largest and this ensures rather conservative estimates.}
\begin{equation}
\left\langle \gamma_\phi \right\rangle_\theta (t):=
\frac
{\int B^\phi (B^r B_r + B^\theta B_\theta)^{-1/2}\sqrt{-g}\,d\theta}
{\int \sqrt{-g}\,d\theta}\,,
\end{equation}
where for symmetry reasons the average is restricted to the upper
hemisphere, \ie $0<\theta < \pi/2$, and to densities in the range
$\rho/\rho_{\mathrm{max}}>10^{-2}$.

\begin{figure}
\includegraphics[width=1.0\columnwidth]{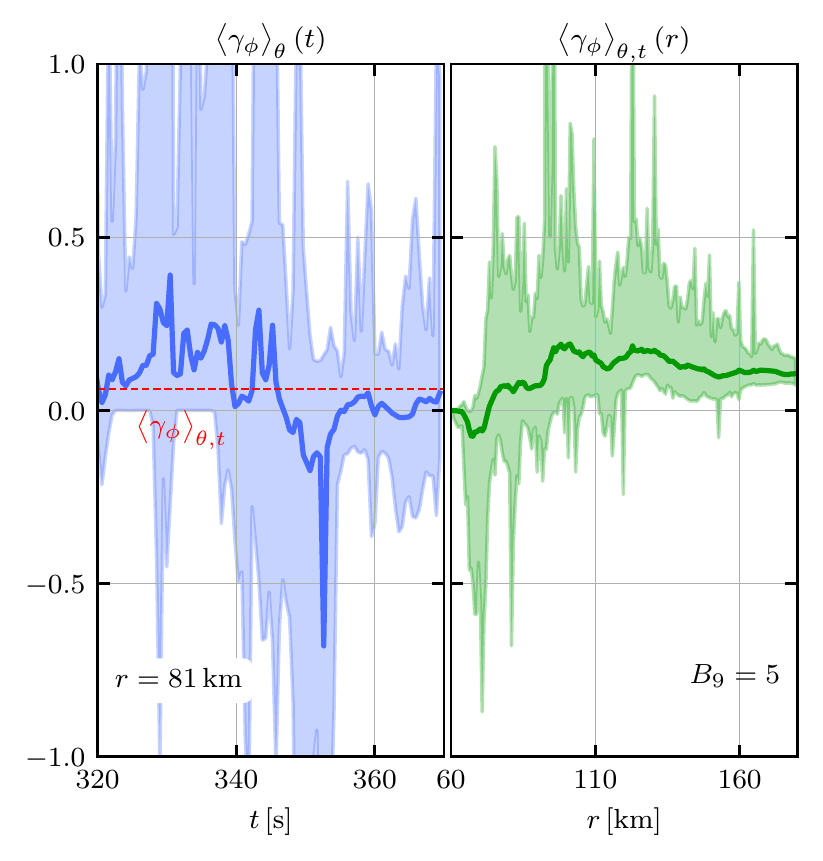}
\caption{Evolution of the pitch factor (left panel) and radial profile of
  the time-averaged pitch factor (right panel). As in previous figures,
  the shaded region marks the maximum and the minimum values of the pitch
  factor measured on a radial shell at $81\,{\rm km}$ and over a
  timescale of $50\,\mathrm{ms}$; the red-dashed line denotes the
  time-averaged value. For symmetry reasons, the polar average is
  restricted to the upper hemisphere, \ie $0<\theta < \pi/2$.
   \label{fig:gamma}}
\end{figure}

Figure~\ref{fig:gamma} reports in its left panel the evolution of the
pitch factor over $50\,\mathrm{ms}$, together with the time-averaged
value marked with a red dashed line. Note that the pitch factor is
essentially zero before a steady-state accretion is established and that
it fluctuates significantly in value, sometimes changing sign as a result
of the turbulent nature of the accretion flow; furthermore, the
fluctuations in time are significantly larger than its average
value. Overall, when averaged over the time window shown in the left
panel of Figure \ref{fig:gamma}, the pitch factor is positive with a
value $\left\langle \gamma_\phi \right\rangle_{\theta,t} \simeq 0.06$.

The right panel of Fig. \ref{fig:gamma}, on the other hand, reports the
radial profile of the time-averaged pitch factor (right). As in previous
figures, the shaded area marks the maximum and the minimum values of the
pitch factor at a given radius over the last $50\,\mathrm{ms}$.
Alternative definitions of the pitch factor -- \eg computed
in terms of the magnetic fields measured in the frame comoving with the
fluid -- do not change
the qualitative behaviour of the pitch factor reported in
Fig.~\ref{fig:gamma}, but do change its range, which can vary of two
orders of magnitude when applied to our simulations 
\citep[see][and references therein]{Hawley2011}.

Simple models, as those proposed by \citet{Ghosh1979a, Campbell1992,
  Wang1995}, have suggested a simple relation between the pitch factor
and the angular velocity of the accreting flow $\Omega$, namely, that
$\gamma_\phi \propto \left(\Omega_*-\Omega\right)/\Omega_K$. As result,
the pitch factor should undergo a sign change at the corotation radius,
where the angular velocity of the plasma equals to the spin of the
star. Because our simulations consider nonrotating stars, we are unable
to verify this prediction, but we can nevertheless explore the relation
between the pitch factor and the angular velocity of the accreting flow.

\begin{figure}
\center \includegraphics[]{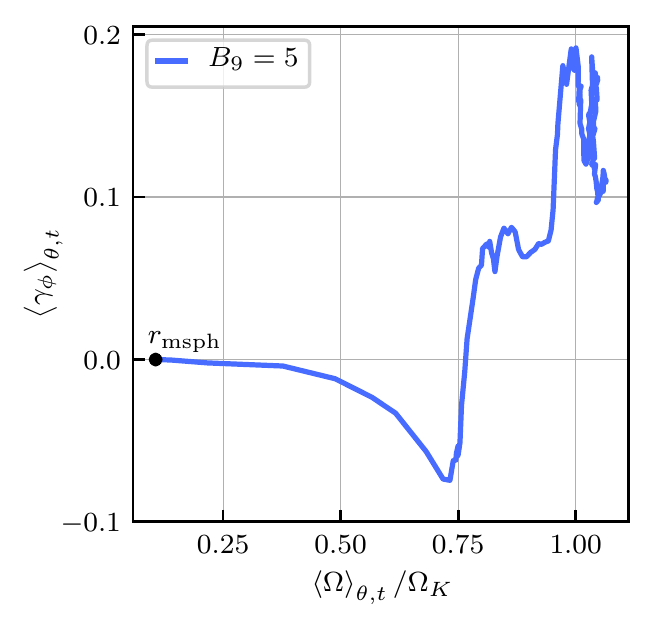}
\caption{The pitch factor as a function of the angular velocity scaled by
  the Keplerian angular velocity. Both quantities are averaged over the
  time (last $50\,{\rm ms}$) and $\theta$ (between $0$ and $\pi/2$).}
\label{fig:gammaOmega}
\end{figure}

This is done in Fig.~\ref{fig:gammaOmega}, which shows the relation
between the polar-averaged pitch factor and the polar-averaged angular
velocity, where the latter is normalised to the Keplerian angular
velocity, \ie $\langle\Omega\rangle_{\theta,t}/\Omega_K$. Because the
accretion flow is far from being laminar near the magnetospheric radius
and the angular velocity of the accreting plasma becomes Keplerian only
at sufficiently large distances (see Fig.~\ref{fig:Omega}), we report in
Fig.~\ref{fig:gammaOmega} the pitch factor between an inner radius
coinciding with the magnetospheric radius $r_{\rm msph}=60\,{\rm km}$
(marked with a black filled circle) and up to an outer radius at
$140\,{\rm km}$.

Clearly, this relation should be a constant according to the analytic
models \citep{Ghosh1979a, Campbell1992, Wang1995}, but this is not what
the simulations actually reveal. This is expected since the ideal-MHD
approximation in our setup does not allow the stellar magnetic field to
penetrate the disc, in contrast with what assumed in the models mentioned
above.  In particular, for values of $\langle \Omega
\rangle_{\theta,t}/\Omega_K \lesssim 0.75$, which corresponds to the
inner parts of the accretion flow and essentially from the local maximum
of the angular velocity in Fig.~\ref{fig:Omega} down to the
magnetospheric radius, the pitch factor is smoothly correlated with
$\langle\Omega\rangle_{\theta,t}/\Omega_K$. It is zero at the
magnetospheric radius and becomes increasingly negative when moving
outwards, reaching a minimum of $\langle\gamma_{\phi}\rangle_{\theta,t}
\simeq -0.1$ for $\langle\Omega\rangle_{\theta,t}/\Omega_K \simeq
0.75$. This is not surprising as in these regions of the flow the angular
velocity is highly turbulent and rather different from the Keplerian
one. However, for $\langle\Omega\rangle_{\theta,t}/\Omega_K \gtrsim
0.75$, the pitch factor rapidly changes sign, becoming positive and
reaches values of the order of $\langle\gamma\rangle_{\theta,t} \simeq
0.2$ in the bulk of the flow. Finally, we note that we find a systematic
and long-lasting sign change in the radial profile of the pitch factor
(see right panel of Fig.~\ref{fig:gamma}) even though this is not
contemplated in the stationary analytic models; this clearly points out
to the inability of such models to capture the highly dynamical flows
encountered in our simulations.

We conclude this section by exploring the properties of the turbulence in
our simulations that, as discussed earlier, is generated by the
development of the MRI. In the classical Shakura-Sunyaev discs, on the
other hand, the molecular turbulent viscosity is expressed in terms of
the shear-viscosity coefficient $\nu := \tilde{\alpha} c_{\rm s} H$
\citep{Shakura1973}, where $c_{\rm s}$ is the speed of sound, $H$ is the
vertical scale-height of the disc, and $\tilde{\alpha}$ -- also referred
to as the ``alpha-viscosity'' parameter -- is an unknown dimensionless
coefficient to be determined by the observations. Typical values for the
dimensionless constant needed to reproduce to some extent the
astronomical observations are $\tilde{\alpha} \simeq 0.1-0.4$ \citep[see,
  \eg][]{King2007, Martin2019}.

In our simulations, we can associate and measure the alpha-viscosity
parameter in terms of the ratio of the Maxwell stresses and of the total
pressure as measured in the frame comoving with the fluid, namely, we
define the polar-averaged MRI-driven alpha-viscosity parameter
$\langle\tilde{\alpha}\rangle_\theta$ \citep{Pessah2006, Shafee2008,
  Porth2019} as
\begin{equation}
\langle\tilde{\alpha}\rangle_\theta := \frac{\int (-b^\phi b^r
  \sqrt{\gamma_{rr}\gamma_{\phi\phi}})\sqrt{-g}\,d\theta}{\int
  (p+b^2/2)\sqrt{-g}\,d\theta} \,,
\end{equation}
where the integral is restricted to to densities in the range
$\rho/\rho_{\mathrm{max}}>10^{-2}$.

\begin{figure}
\includegraphics[width=1.0\columnwidth]{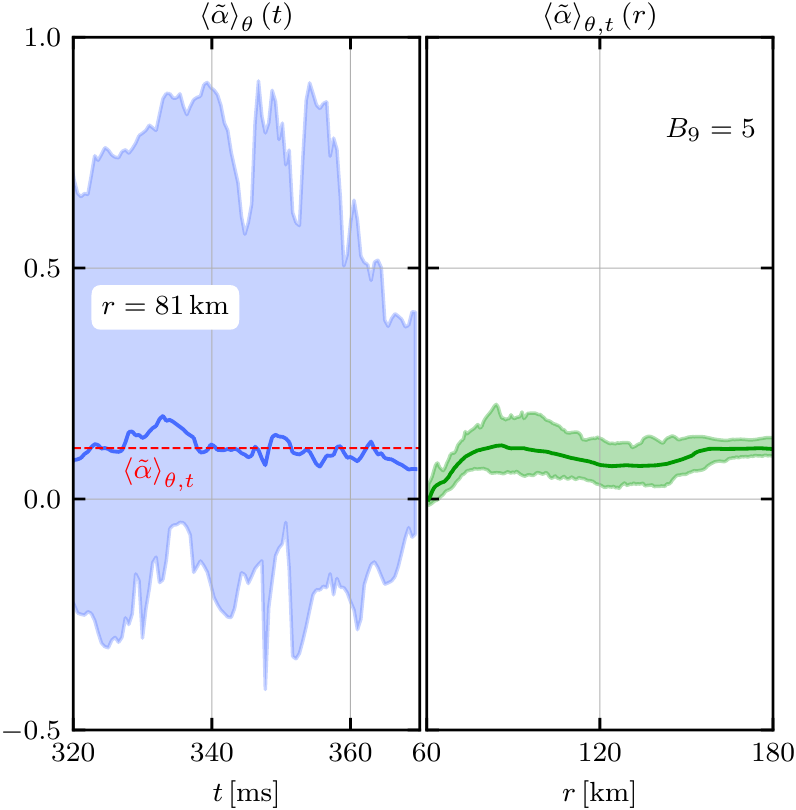}
\caption{Time evolution of the
  $\langle\tilde{\alpha}\rangle_\theta$-viscosity parameter at
  $81\,{\mathrm{km}}$ (left) and radial profile of the time averaged
  $\langle\tilde{\alpha}\rangle_\theta$-viscosity parameter over the last
  $50\,\mathrm{ms}$ (right) for $B_9=5$ case. As in previous figures, the
  shaded area marks the maximum and the minimum values of the
  $\langle\tilde{\alpha}\rangle_\theta$-viscosity parameter at a given
  radius over the last $50\,\mathrm{ms}$, and the red-dashed line denotes
  the time-averaged value.
    \label{fig:alpha}}
\end{figure}

We report in Fig.~\ref{fig:alpha} the evolution of
$\langle\alpha\rangle_\theta$-viscosity parameter at a fixed radius of
$81\,{\rm km}$ as well as with its variance in space (left panel), and
its radial profile averaged over the last $50\,{\rm ms}$, together with
its fluctuations (right panel). Both panels refer to the $B_9=5$ model
and indicate that the time-averaged value of the MRI-driven alpha
viscosity is $\langle\alpha\rangle_\theta=0.11$ at $r=81\,{\mathrm{km}}$.

Clearly, our simulations reveal that
$\langle\tilde{\alpha}\rangle_\theta$ fluctuates significantly both in
space and time. While this was already suggested by
\citet{Lyubarskii1997}, it points out to important differences between an
MRI-driven alpha viscosity -- which is unsteady in space and time -- and
the alpha-viscosity used in the simulations of \citet{Romanova2002} --
which is instead constant in space and time. On the other hand, similar
results have been presented by \citet{Romanova2011}, who report an
MRI-driven alpha viscosity in range of $\langle\alpha\rangle_\theta
\simeq 0.01-0.1$.

\section{Discussion and Conclusion}
\label{sec:Dis}

We have reported 2D general-relativistic magnetohydrodynamics (GRMHD)
simulations of matter being accreted onto and ejected by a magnetised and
nonrotating neutron star. The dynamics is followed within the ideal-MHD
limit and making use of the numerical code \texttt{BHAC}. 

Employing a number of simulations and considering various strengths of
the stellar dipolar magnetic field, we have determined self-consistently
the location of the magnetospheric radius $r_{\rm msph}$ and study how it
depends on the magnetic moment $\mu$ and on the accretion rate. Overall,
we recover the analytic Newtonian scaling relation, \ie $r_{\rm msph}
\propto B^{4/7}$, confirming the behaviour explored by
\citet{Parfrey2017}.  At the same time, we find that the correlation with
the mass-accretion rate is different from the Newtonian expectation, \ie
$r_{\rm msph} \propto \dot{m}_{\mathrm{in}}^{-2/7}$, and recently
observed in the Newtonian simulations of \citet{Ireland2022}. The weaker
correlation with the accretion rate found here, \ie $r_{\rm msph} \propto
\dot{m}_{\mathrm{in}}^{-1/12}$, is unlikely to be due to our
general-relativistic framework, which we show in
Appendix~\ref{sec:MagRad} to provide only second-order corrections to the
magnetospheric radius. Rather, we believe that the turbulent nature of
the accreting flow produced in our simulations weakens the dependence of
the magnetospheric radius on the accretion rate, leaving the magnetic
field as the main regulator of its location.

As is natural to expect, the fluctuations in the mass-accretion rates are
accompanied by the fluctuations in the matter-part (\ie independent of
the magnetic-field strength) of the angular-momentum transport rate. Our
simulations exhibit a clear correlation between the mass-accretion rate
and the matter-part of the rate of transport of angular momentum. 
This correlation supports the idea that the episodic reconnection
processes taking place at the edge of the magnetosphere should lead both
to an increased luminosity and to a stellar spin-up. Interestingly, when
expressing this correlation as $\dot{j}_{\mathrm{matt, in}}\propto
\dot{m}_{\mathrm{in}}^{\lambda}$, we can estimate the exponent of the
correlation to be $\langle \lambda \rangle=1.18$, which is in good
agreement with the analysis of 12 X-ray sources reported by
\citet{Sugizaki2017}.
  
We note that in our simulations the total angular-momentum transport rate
exhibits large fluctuations that become larger as the magnetic field of
the star is increased. These fluctuations are due to the complex plasma
dynamics taking place at the magnetospheric radius and can be so intense
that can lead to a change of sign in the total angular-momentum transport
rate and could therefore lead to a spin-down of the accreting star.
Although these sign changes occur over timescales that are much shorter
than those measured in the observations, the phenomenology produced in
our simulations may have implications on the torque reversals or on the
noise in the measured spin frequencies in the case of slowly rotating
stars. In addition, we have investigated the behaviour of the pitch
factor and found it to be quite different from what is expected in
simplified models of magnetically-threaded discs. More specifically, we
have observed that the pitch factor undergoes significant fluctuations in
time and space -- sometimes undergoing sign changes -- as a result of the
turbulent nature of the accreting flow. These differences are not
surprising given the ideal-MHD assumption under which our simulations are
carried out, which prevents the stellar magnetic field from penetrating
the accreting plasma. Yet, the significant differences -- both
qualitative and quantitative -- found in the properties of the pitch
factor casts doubts on the effectiveness of this diagnostic quantity in
characterising the properties of the accreting flow.

Finally, our results confirm the findings of \citet{Romanova2011} in that
discs with MRI-driven turbulence have substantial differences when
compared to discs with constant ${\alpha}$-viscosity parameter. However,
when expressing this parameter in terms of the ratio of the Maxwell
stresses and of the total pressure as measured in the frame comoving with
the fluid, it exhibits large fluctuations both in space and in time, but
these average to values of ${\alpha} \sim 0.1$.

Future work will improve on several aspects of these
simulations. First, while the numerical resistivity of our ideal-MHD
simulations is very small and unable to change the bulk properties of
the accretion flow (\eg the diffusion of the stellar magnetic field
lines in the accretion disc), it is important to assess how resistive
effects impact on the dynamics described here and on the scaling
relations found. To this scope, we will introduce a physical model of
resistivity along the lines of similar simulations carried out by
\citet{Ripperda2020} and \citet{Nathanail2021b}. This will allow us to
investigate under more realistic conditions the generation of the
toroidal magnetic field and its effect on the torque exerted onto the
star.  Second, the 2D nature of our simulations has the consequence
that turbulence and magnetic fields intrinsically decay unless a proper
dynamo mechanism is implemented \citep{Sadowski2015}, and the
non-axisymmetric motion of the fluid that can create differences in the
dynamics of the accretion \citep{Igumenshchev2008, Romanova2012,
McKinney2012} is ignored. Hence, we will reconsider the ideal-MHD
scenarios explored here also in fully three-dimensional (3D)
simulations. Finally, to better investigate the interaction between the
magnetosphere and the accretion disc, our investigations (either in 2D
or in 3D) will be extended by considering the impact of stellar
rotation.

\medskip
While completing this work, we have become aware of a similar and
independent study of \citet{Das2022}, which investigates accretion onto
rotating stars with a dipolar and quadrupolar magnetic-field
topologies. While the numerical code employed is the same
\citep{Porth2017,Olivares2019}, the scope of the paper is different and
differently from us, \citet{Das2022} introduce a force-free prescription
in the magnetosphere necessary to handle a rotating star. Despite these
differences, the phenomenology observed under the same physical
conditions (\ie dipolar magnetic field, comparable magnetisation) is
similar, as is the location and scaling of the magnetospheric radii.

\section*{Acknowledgements}

LR gratefully acknowledge funding by the State of Hesse within the
Research Cluster ELEMENTS (Project ID 500/10.006), by the ERC Advanced
Grant ``JETSET: Launching, propagation and emission of relativistic jets
from binary mergers and across mass scales'' (Grant No. 884631). The
numerical calculations reported in this paper were performed on Iboga
Cluster in Frankfurt, and at T\"UB\.ITAK ULAKBIM, High Performance and Grid
Computing Center (TRUBA resources) in Ankara. KYE acknowledges support 
from the Scientific and Technological Research Council of Turkey
(T\"UB\.ITAK) with project number 112T105. S\c{C} acknowledges support
from T\"UB\.ITAK with grant number 1059B141801188.

\section*{Data Availability}
The data underlying this article will be shared on reasonable request
to the corresponding author.



\bibliographystyle{mnras}
\bibliography{aeireferences} 


\appendix
\section{Magnetospheric Radius}
\label{sec:MagRad}

We here present an analytical estimate of the magnetospheric radius in a
general-relativistic framework by using the balance of pressures.
Because the magnetospheric radius is located at a few times the radius of
the star, we treat relativistic effects as small corrections, $M/r\ll 1$,
and retain terms at the second-order in the expansion
$\mathcal{O}(M/r)$. In this framework, the condition of pressure balance
between the total and the magnetic pressures can be written as
\begin{equation}
\rho h \Gamma^2 v^2+p=\frac{b^2}{2}\, \label{eq:rmag_condition},
\end{equation}
Adopting an ideal-fluid equation of state for the left-hand side of
\eqref{eq:rmag_condition} yields
\begin{equation}
\rho h \Gamma^2 v^2+p=\left(\rho + \rho \epsilon + p\right)\Gamma^2v^2 +
p = \left(\rho + p\frac{\gamma}{\gamma-1}\right)\Gamma^2 v^2+p\,,
\end{equation}
where the last term in the equation above will be neglected hereafter
since in the innermost regions of torus the pressure is much smaller than
the rest-mass density. Assuming now, and for simplicity, a spherically
symmetric accretion flow in steady state with $v^r=v$, the corresponding
mass-accretion rate can be written as
\begin{equation}
\dot{m}=4\pi\rho r^2 \Gamma v \,,
\end{equation}
and the relativistic velocity $\Gamma v$ be expressed as
\begin{equation}
\Gamma^2 v^2= g_{rr}\Gamma^2 \left(v_{\mathrm{ff}}\right)^2
=\left(1-\frac{2M}{r}\right)^{-1}\frac{2M}{r}\,, 
\end{equation}
so that the rest-mass density can be eliminated in
Eq.~\eqref{eq:rmag_condition}. Moreover, it is reasonable to assume that
magnetic field at the magnetospheric radius is essentially the stellar
one, so that on the equatorial plane, $\theta=\pi/2$, this is given by
\begin{align}
B^r =&\, 0, \\
B^\theta =&\, \frac{\mu}{r^4}\sqrt{1-\frac{2M}{r}}\left[ \frac{3r^3}{4M^3}\ln\left(1-\frac{2M}{r}\right)+\frac{3r^2}{4M^2}\frac{2r-2M}{r-2M}\right]
\notag \\
\simeq& \,\frac{\mu}{r^4}\left[1+\frac{2M}{r}+\frac{37M^2}{10r^2}+\mathcal{O}\left(\frac{M^3}{r^3}\right)\right]\,,
\label{eq:bp}
\end{align}
up to the second-order correction in an expansion in terms of $M/r$
\citep{Wasserman1983,Rezzolla2001}.
Finally, the strength of the magnetic field in the fluid frame at the
equatorial plane is given by
\begin{align}
b^2 =&\, \frac{B^2}{\Gamma^2}+\left(B^i v_i\right)^2
=r^2\left(B^\theta\right)^2\left(1-\frac{2M}{r}\right) \notag
\\ \simeq&\, \frac{\mu^2}{r^6}\left[1+\frac{2M}{r}+\frac{17M^2}{5r^2} +
  \mathcal{O}\left(\frac{M^3}{r^3}\right)\right]\,.
\label{eq:b2}
\end{align}

Inserting \eqref{eq:bp} and \eqref{eq:b2} in
Eq.~\eqref{eq:rmag_condition}, we obtain that the magnetospheric radius
at the second order in the $\mathcal{O}(M/r)$ expansion can be
approximated as
\begin{equation}
  \label{eq:rmsph_gr_appendix}
  r_{\mathrm{msph}}^{7/2}\left(1+\frac{3M^2}{5r_{\mathrm{msph}}^2}\right)
  \simeq \frac{\mu^2}{\sqrt{8M}\dot{m}}\,.
\end{equation}
%


\bsp	
\label{lastpage}
\end{document}